%% file: paper.tex
\documentclass[aps,prd,superscriptaddress,floatfix,nofootinbib,preprintnumbers]{revtex4}
\usepackage{amsmath}
\usepackage{graphicx,tabularx}
\usepackage{url}
\usepackage{footmisc}

\newcommand{\ie}{{\sl i.e. }}
\newcommand{\eg}{{\sl e.g. }}

\include{declare}

\begin{document}

\date{\today}

\title{W+Jets at CDF: Evidence for Top Quarks}

\author{Tilman Plehn}
\affiliation{Institut f\"ur Theoretische Physik, Universit\"at Heidelberg, Germany}

\author{Michihisa Takeuchi}
\affiliation{Institut f\"ur Theoretische Physik, Universit\"at Heidelberg, Germany}

\begin{abstract}
  Recently, an anomaly of $W$+jets events at large invariant masses
  has been reported by CDF. Many interpretations as physics beyond the
  Standard Model are being offered. We show how such an invariant mass
  peak can arise from a slight shift in the relative normalization of
  the top and $WW$ backgrounds.
\end{abstract}

\maketitle


In recent years, the Tevatron experiments have run a successful search
program studying weak gauge boson and top quarks we well as searching
for a Higgs boson and for new physics. A specific search for $V$+jets
production $(V=W,Z)$~\cite{cdf_ww,cdf_tc,theo_wjets} follows a long
list of motivations~\cite{review}: we can test QCD effects such as the
so-called staircase scaling of $n$-jet production~\cite{theo_scaling},
we can search for triple gauge boson couplings in $W^+ W^-$ and $W^\pm
Z$ production~\cite{cdf_ww,theo_ww}, we can search for technicolor
signals~\cite{exis_tc,theo_tc}, or in the case of two bottom jets we
can look for $WH$ associated production~\cite{exis_wh}. Some
of these channels include a study of the invariant mass of the two
leading jets recoiling against a leptonically decaying $W$ boson
\begin{equation}
 p \bar{p} \to (W \to \ell \nu) \, + \text{2 jets} \, + X
 \qquad \qquad \qquad \qquad
 (\ell = \mu, e)
\end{equation}
In their published study of $WV$ production based on an integrated
luminosity of $3.9~\ifb$ and focused on an invariant mass regime
$m_{jj} = 50 - 130$~GeV the CDF collaboration has started to observe a
slight excess of events in the region of $m_{jj} =
150-180$~GeV~\cite{cdf_ww}. A D0 search based on the lower luminosity
of $1.1~\ifb$ does not show any excess in this mass
range~\cite{d0_ww}.

More recently, the CDF collaboration has published a dedicated study
of the same anomaly~\cite{cdf_tc} with harder background rejection
cuts and reports a $3.2~\sigma$ anomaly in the $m_{jj}$ spectrum. The
excess is compatible with a resonance around 150~GeV. Many papers have
since been published, explaining this observation, including
technicolor~\cite{bsm_tc}, supersymmetric~\cite{bsm_susy},
lepto-phobic $Z'$ boson~\cite{bsm_zprime}, color
octets~\cite{bsm_octet}, and other
interpretations~\cite{bsm_other}. In this paper we suggest an
explanation of the excess based on a slight relative shift of the
weight of different background contributions on the $WV$ pole an in
the higher-mass region. While it is certainly possible to relieve the
tension of the measurement and the background prediction for example
by a shift in $m_{jj}$ or through the heavy flavor content of the
proton, to our knowledge ours is the only way to explain the observed
kinematic feature within the Standard Model.

\subsection*{A second peak from top decays}
\label{sec:top}

One of the backgrounds to $W$+jets production is the production of top
quarks. Unlike to all other Standard Model channels, top quarks lead
to a second peak in the $m_{jj}$ distribution, in addition to the $W$
mass peak. The angular correlation behind this second peak is between
the bottom and the up-type quark $q_{\uparrow}$ from the $W$-decay. In
the $W$ rest frame the distribution is given by
\begin{equation}
P(\cos\theta) 
= \frac{3}{8} \, \left( 1+\cos\theta \right)^2 \; F_R 
+ \frac{3}{8} \, \left( 1-\cos\theta \right)^2 \; F_L 
+ \frac{3}{4} \, \sin^2\theta F_0 \; .
\end{equation}
In the Standard Model the relative size of these contributions is
$F_0:F_L:F_R \simeq 0.7:0.3:0$~\cite{top_kin}. The corresponding
invariant mass $m_{bq_{\uparrow}}$ is
\begin{equation}
P(m_{bq_{\uparrow}}) = \frac{f_R(r) F_R +  f_L(r)  F_L + f_0(r) F_0 }{m_{bj}^{\max}} 
\qquad \qquad \text{with} \quad
r= \frac{m_{bq_{\uparrow}}}{m_{bj}^\text{max}}= \sqrt{\frac{1-\cos\theta}{2}} \; ,
\end{equation}
$f_R(r)=6r(1-r^2)^2$, $f_L(r)=6r^5$, and $f_0(r)=12 r^3(1-r^2)$.
Its upper endpoint is $m_{bj}^\text{max} =\sqrt{m_t^2 -
  m_W^2}=154.6$~GeV, neglecting the bottom mass.\medskip  

\begin{figure*}[!t]
  \includegraphics[width=0.4\textwidth]{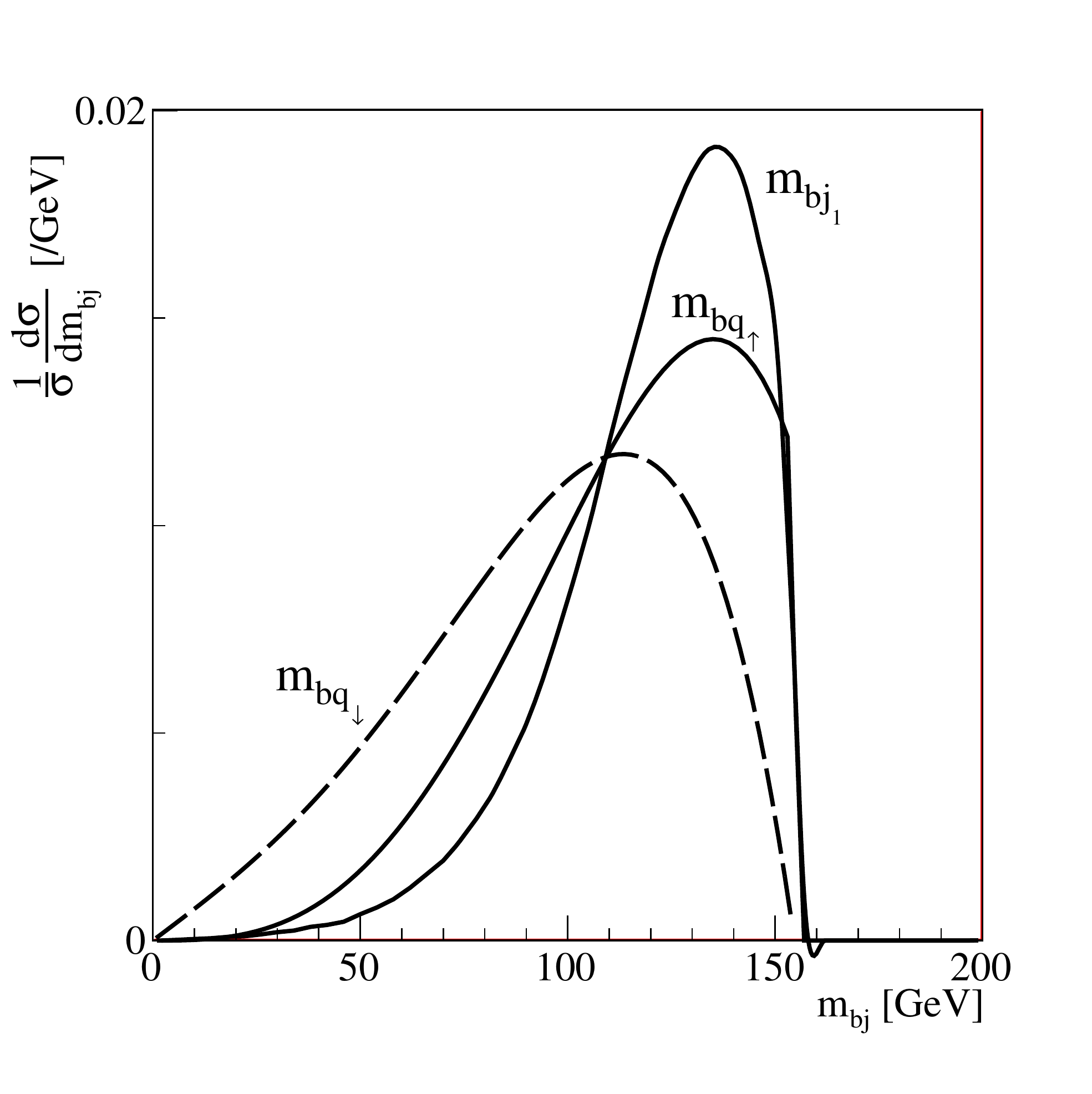}
    \hspace*{0.1\textwidth}
  \includegraphics[width=0.4\textwidth]{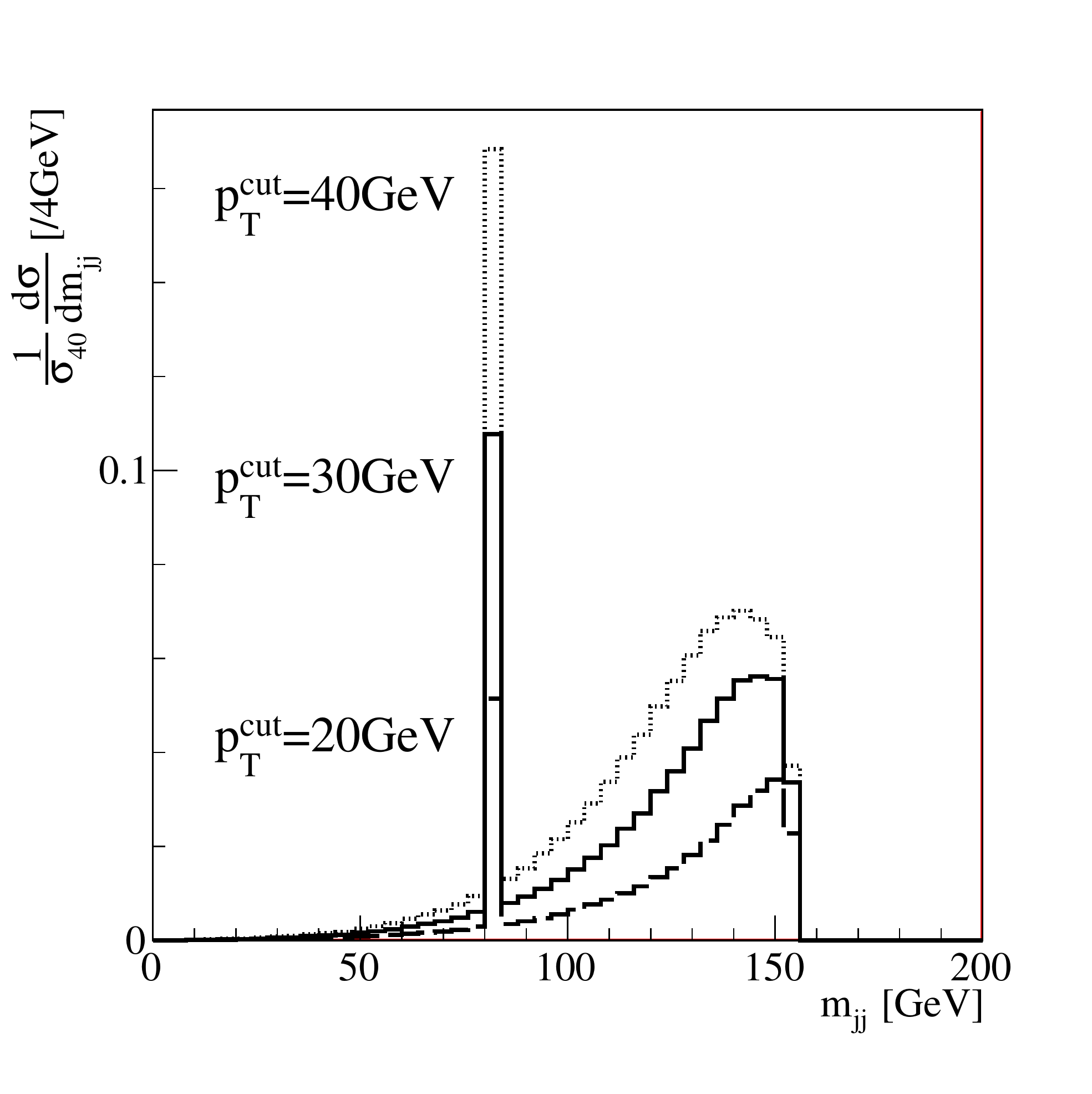}
\vspace*{-5mm}
  \caption{Left: theory predictions for the $m_{bq_{\uparrow}}$,
    $m_{bq_{\downarrow}}$ and $m_{bj_1}$ distributions at the parton
    level, simulated for top pair production.  Right: $m_{j_1 j_2}$
    distribution with $p_{T,{j_3}} < 20$~GeV (dashed), 30~GeV (solid)
    and 40~GeV (dotted).}
 \label{fig:peaks}
\end{figure*}

The theory prediction for the $m_{bq_{\uparrow}}$ distribution we show
in Fig.~\ref{fig:peaks}. Because of the left-handed $W$ interaction
$m_{bq_{\uparrow}}$ gets contributions from $f_L$ and $f_0$;
$m_{bq_{\downarrow}}$ corresponds to exchanging $f_L$ and $f_R$.
Experimentally, we cannot distinguish between $q_{\uparrow}$ and
$q_{\downarrow}$, so instead we define the invariant mass $m_{b j_1}$
with the harder of the two $W$ decay jets.  This distribution is
harder than $m_{bq_{\uparrow}}$.  Without $b$ tagging the only
observable distribution is $m_{j_1 j_2}$, using the hardest two jets
from the top decay. It shows a double peak structure from the sum of
the $W$ peak and the $m_{bj_1}$ distribution. In Fig.~\ref{fig:peaks}
we also show how a stricter jet veto not only reduces the number of
events but also produces a harder second peak in $m_{jj}$.

\subsection*{Loose cuts}
\label{sec:loose}

In this first part of our paper we look at the original $WV$ analysis
with the less significant but nevertheless clearly visible excess,
shown in the left panel of
Fig.~\ref{fig:shapes_loose}~\cite{cdf_ww,thesis}. The basic acceptance
and background rejection cuts are on one lepton and at least two jets
plus missing transverse energy with
\begin{alignat}{7}
E_{T,\ell} &> 20~\gev  \qquad \qquad 
& \met &< 25~\gev  \qquad \qquad 
& M_{T,W} &> 30~\gev \notag \\[2mm]
E_{T,j} &> 20~\gev  \qquad \qquad 
& |\eta_j| &< 2.4   \qquad \qquad 
& |\Delta \phi_{\met, j_1}| &>0.4 \notag \\
p_{T,jj} &> 40~\gev  \qquad \qquad 
& |\Delta \eta_{jj}| &< 2.5  \; .
\label{eq:cuts1}
\end{alignat}
The main background is $W$+jets production with a variable
normalization which can be fixed from the shape of the $m_{jj}$
distribution. This background shows essentially no structure. The
second background is QCD jet production faking a lepton and missing
transverse energy. For a $W$ decaying to an electron this background
is about four times the size of the muon decay
signature~\cite{thesis}. Again, this background has no visible
structure in $m_{jj}$. 

Of roughly similar size is the top background, consisting of top pairs
and of single top production. As discussed above, this background has
a distinct shape, namely two peaks including a Jacobian peak around
140~GeV. We see this shape in the right panel of
Fig.~\ref{fig:shapes_loose}. The peak arises if we combine the $b$ jet
with one of the two light-flavor jets from the $W$ decay, which means
it gets contributions from top pair production and from single top
production with a $W$ boson. In the analysis, this background is
normalized to the theory predictions $\sigma_{t\bar{t}} = 7.5$~pb and
$\sigma_{\text{single} \, t}=2.9$~pb~\cite{theo_tops}. The signal in this
analysis is $WV$ production. It has a clear peak dominated by $W^+
W^-$ production at $m_{jj} = 80$~GeV, smeared by the experimental
resolution.  Its extracted rate, corrected to the total cross section
without any detector effects or branching ratios is $13.5 \pm 4.4$~pb
for electrons and $23.5 \pm 4.9$~pb for muons. In combination this
gives $18.1 \pm 3.3(\text{stat}) \pm 2.5(\text{syst})$~pb. This
combined number is compatible with the theory prediction.\medskip

However, the two significantly different results for the electron and
the muon analyses with their different background compositions mostly
in the $Z$+jets and QCD jets channels raise the question how well we
actually know the total composition of all backgrounds. For
backgrounds which do not have a distinct $m_{jj}$ shape this question
is not very relevant, but for the top background and the $WV$ signal
it matters. In the right panel of Fig.~\ref{fig:shapes_loose} we first
show the individual templates for the top background and for the $WV$
channel.  Our simulation is based on {\sc Alpgen}~\cite{alpgen} + {\sc
  Pythia}~\cite{pythia} at the particle level.  To model the measured
$m_{jj}$ distribution we apply a Gaussian smearing. Our template
$m_{jj}$ distributions reproduce the CDF results~\cite{thesis}. The
normalization we fix to the $4.3~\ifb$ of Ref.~\cite{thesis}, to
properly take into account detector effects and efficiencies. This
means that whenever we discuss the normalization of different cross
sections we refer to the total rate after efficiencies and detector
effects.

\begin{figure*}[!t]
  \includegraphics[width=0.40\textwidth]{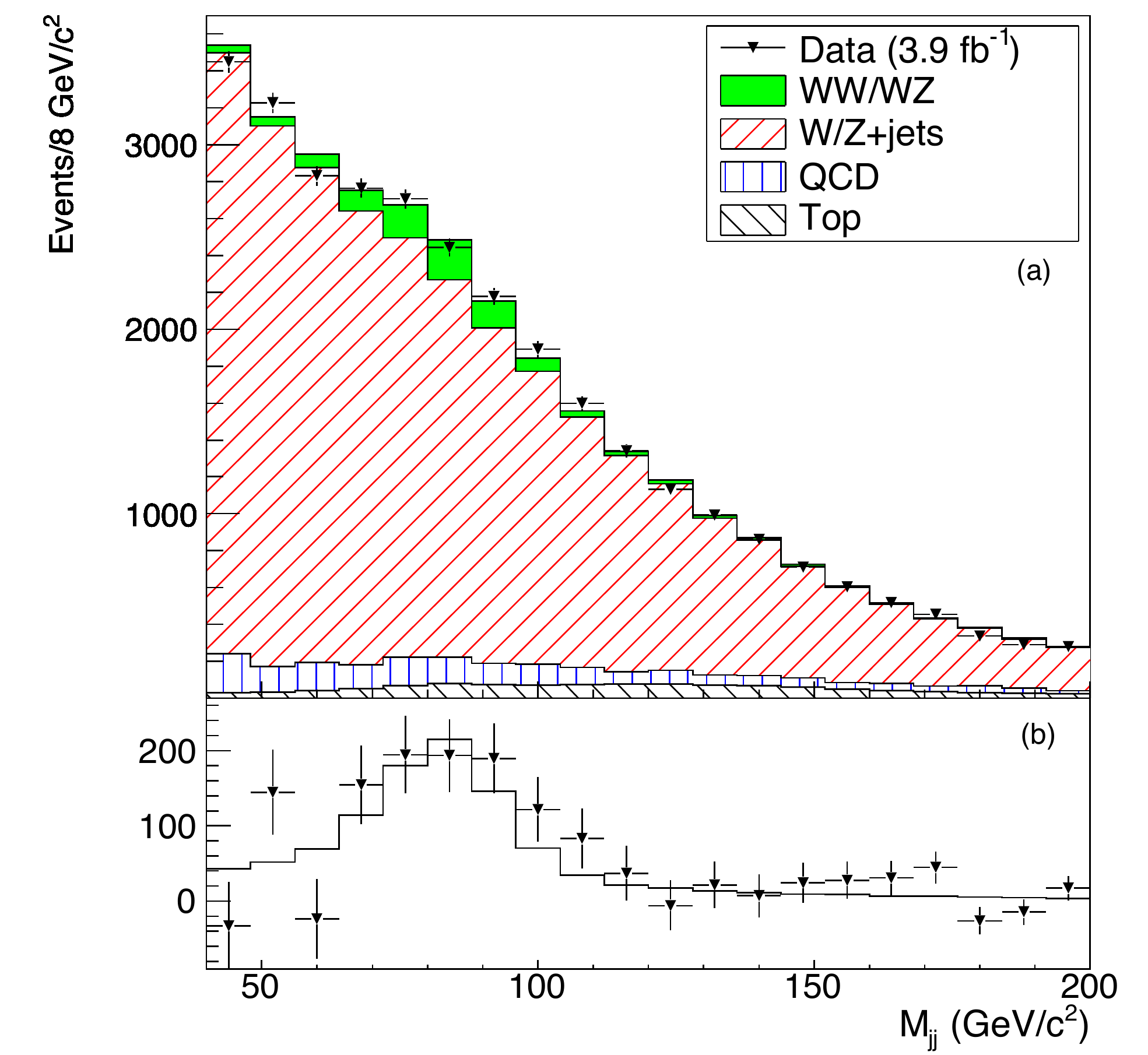}
  \hspace*{0.1\textwidth}
  \includegraphics[width=0.40\textwidth]{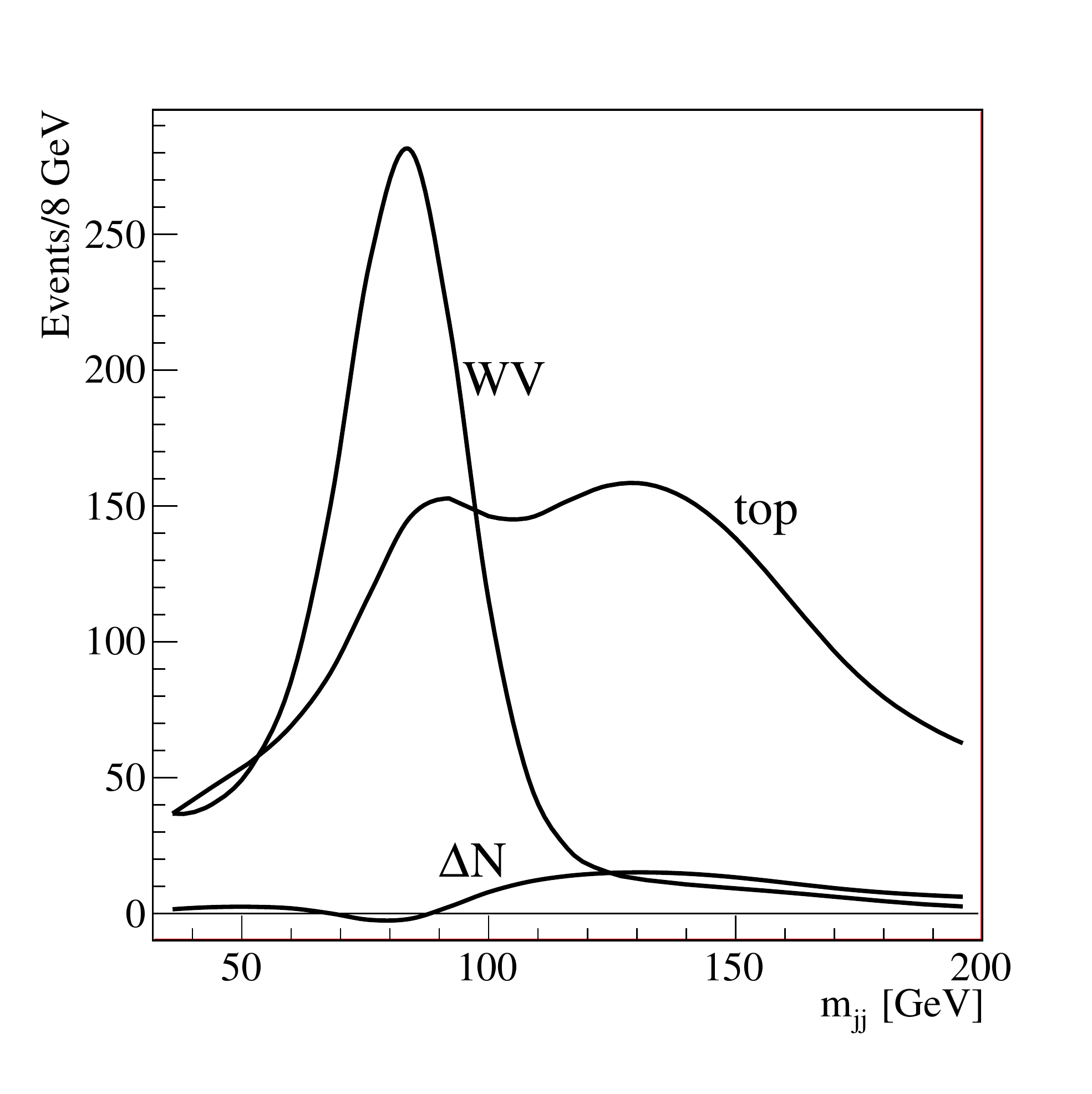}
\vspace*{-2mm}
  \caption{Left: excess in the $m_{jj}$ distribution above the $WV$
    peak, as reported by CDF~\cite{cdf_ww}. The electron and muon
    decay channels are added. Right: $m_{jj}$ templates for the $WV$
    and top samples individually. The normalization is chosen to match
    the CDF data. We also show the difference between the two samples
    for a 10\% change of $\sigma_\text{top}$ and a corresponding shift
    in $\sigma_{WV}$, as described in the text.}
 \label{fig:shapes_loose}
\end{figure*}

The difference between the two templates becomes relevant if we change
the relative contributions of the top and $WV$ backgrounds. The difference
clearly matches the slight observed excess.  To quantify this effect
we compute the change in event numbers associated with a shift of the
integrated rate or efficiency.  We independently consider the peak
region and the high mass regime
\begin{alignat}{5}
\Delta N_{[64,96]}  
&=   \; 926 \; \frac{\Delta \sigma_{WV} }{\sigma_{WV} }
&+&  \; 542 \; \frac{\Delta \sigma_\text{top}}{\sigma_\text{top}}
\notag \\
\Delta N_{[120,170]} 
&=   \; 88 \; \frac{\Delta \sigma_{WV} }{ \sigma_{WV} }
&+&  \; 915 \; \frac{\Delta  \sigma_\text{top}}{\sigma_\text{top}} \; .
\label{eq:loose1}
\end{alignat}
These event numbers correspond to the CDF
analysis~\cite{thesis}. Requiring that the sensitive normalization of
the $WV$ mass peak $m_{jj} = 64 - 96$~GeV be unchanged relates the two
shifts as $(\Delta \sigma_{WV})/\sigma_{WV} = -0.59 \, (\Delta
\sigma_\text{top})/\sigma_\text{top}$, assuming efficiencies do not
vastly vary between the two mass windows. Using this relation we find
a net shift in the high mass region
\begin{equation}
\Delta N_{[120,170]} 
= \; 863 \; \frac{\Delta \sigma_\text{top}}{\sigma_\text{top}} \; .
\label{eq:loose2}
\end{equation}
Throughout this paper $\sigma$ really means the cross section after
cuts and efficiencies, \ie $\sigma \times \epsilon_\text{cuts} \times
\epsilon_\text{rec}$.  The shape of the difference we show in
Fig.~\ref{fig:shapes_loose} for $(\Delta
\sigma_\text{top})/\sigma_\text{top} = 10\%$.  The experimentally
observed excess for the loose set of cuts has the same shape.  In
Fig.~\ref{fig:shapes_loose} the mass window $m_{jj} = 120 - 170$~GeV
includes roughly 100 events which usually are attributed to the $WV$
contribution and any kind of new physics. If we conservatively neglect
possible $WV$ contributions, according to Eq.(\ref{eq:loose2}) this
corresponds to an $\mathcal{O}(10\%)$ shift in the combined top
rate.\medskip

For the sum of top pairs and single top production with its different
hard processes this shift could arise from a combination of
experimental efficiencies and distributions mostly of the many jets
involved. For example, the number of events which we expect from the
combined top sample is very sensitive to the $p_{T,j}$ requirements we
apply.  Moreover, from the CDF publications~\cite{cdf_ww,thesis} it is
not clear how exactly the $tW$ single top channel has been
computed~\cite{pdf}. Its size before cuts ranges around 1\% of the top
pair cross section~\cite{single_top}, but after the cuts
Eq.(\ref{eq:cuts1}) it could well account for a larger fraction of the
shift in relative normalization.

The compensating shift in the $WV$ rate is even smaller and clearly
within the sizable uncertainties of up to $\mathcal{O}(30\%)$ for the
individual decay channels. In short, a very slight shift of the top
sample normalization after cuts and efficiencies compensated for by a
shift of the $WV$ rate completely explains the observed high-$m_{jj}$
anomaly.  We should, however, remark that this loose cuts analysis is
not a serious challenge to Standard Model explanations. It only serves
as a way to illustrate and check our approach before we apply it to
the more challenging dedicated analysis~\cite{cdf_tc}.

\subsection*{Hard cuts}
\label{sec:hard}

\begin{figure*}[!t]
  \includegraphics[width=0.40\textwidth]{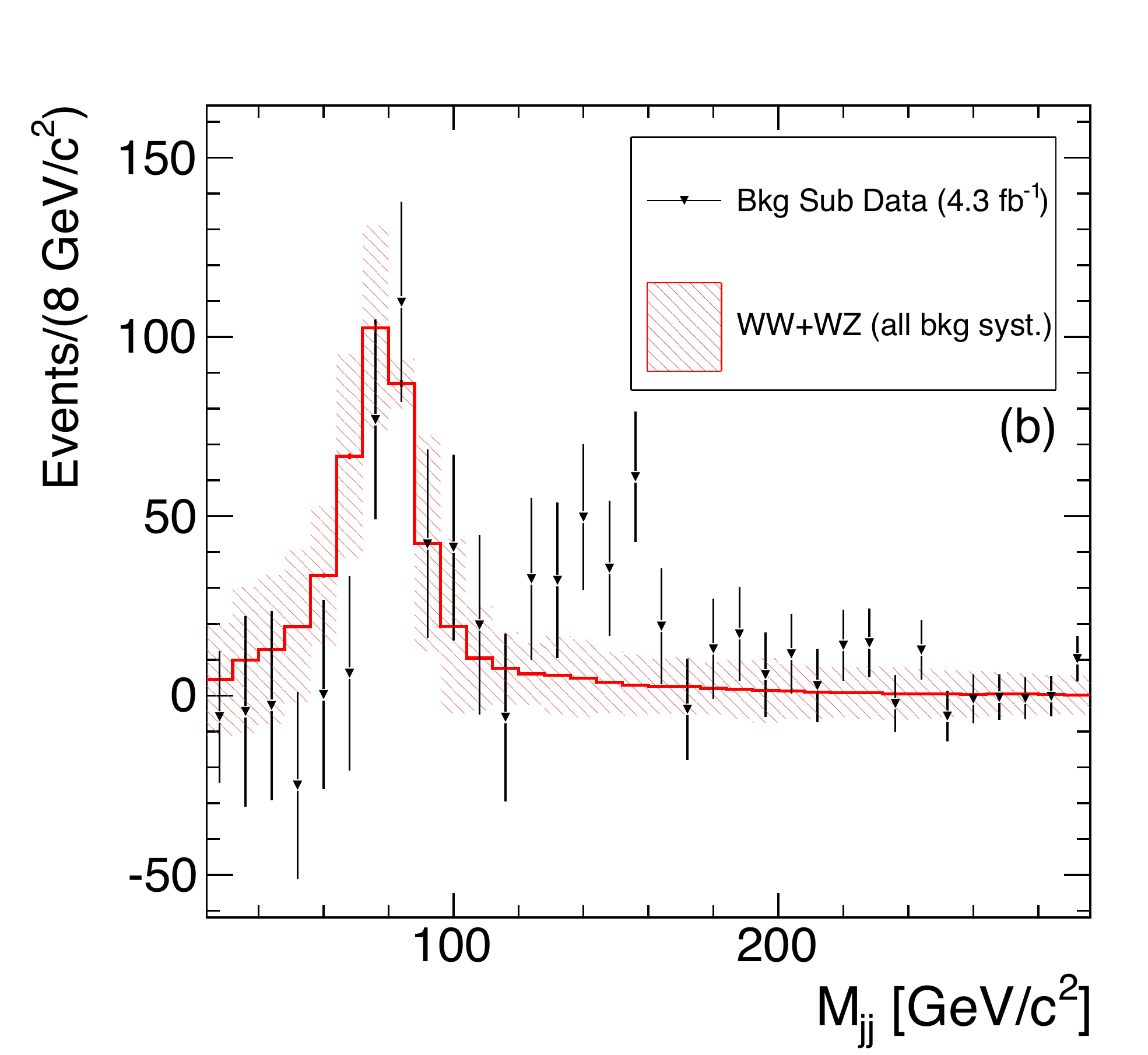}
  \hspace*{0.1\textwidth}
  \includegraphics[width=0.40\textwidth]{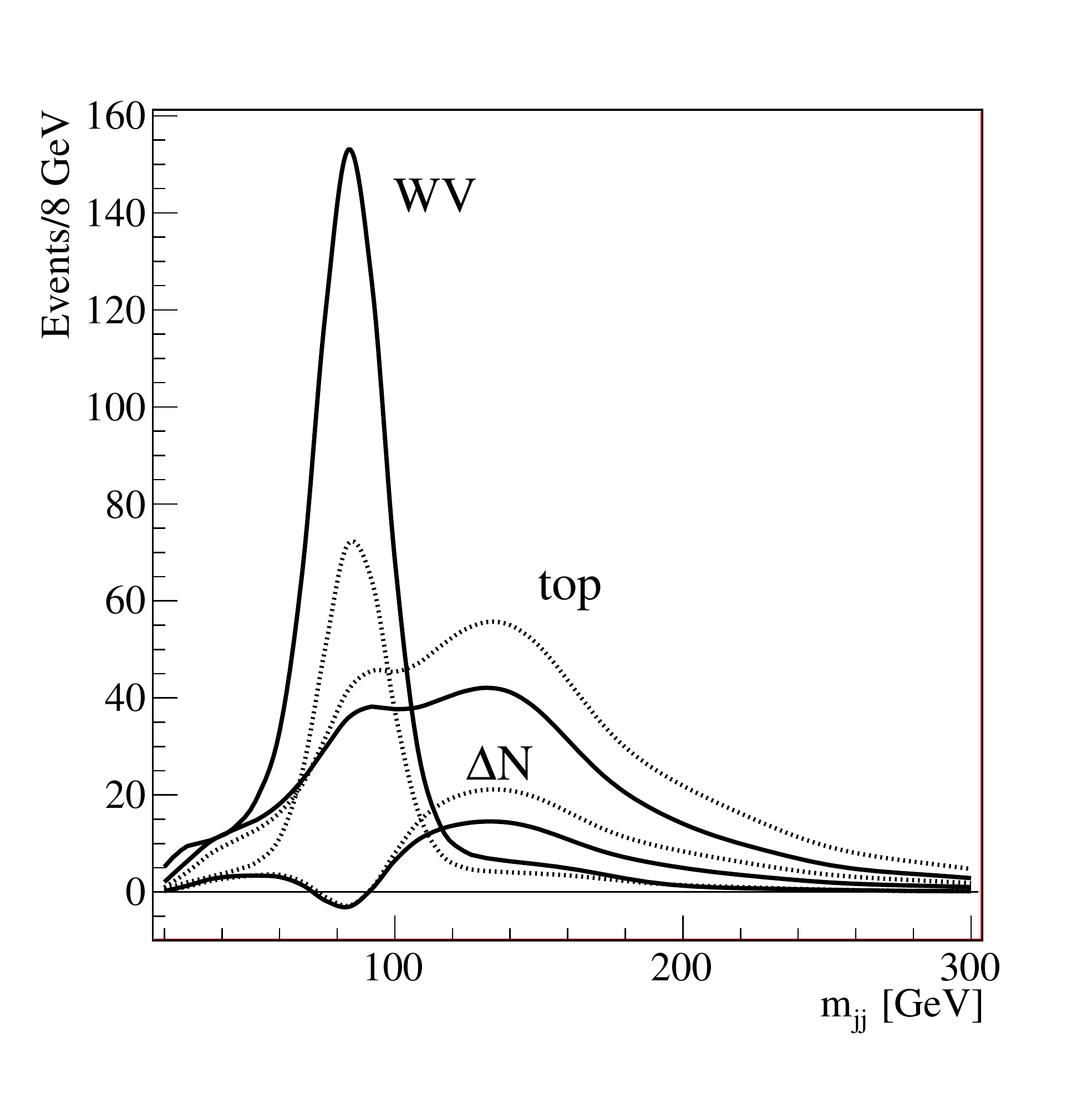}
\vspace*{-2mm}
  \caption{Left: excess in the $m_{jj}$ distribution above the $WV$
    peak, as reported by CDF~\cite{cdf_tc}. The electron and muon
    decay channels are added. Right: $m_{jj}$ templates for the $WV$
    and top samples individually. The dark lines assume
    $E_{T,j}=30$~GeV for the jet criteria, the lighter lines 40~GeV.
    We also show the difference between the two samples for a 40\%
    change of $\sigma_\text{top}$ and a corresponding shift in
    $\sigma_{WV}$, as described in the text.}
\label{fig:shapes_hard}
\end{figure*}

After observing the $m_{jj}$ anomaly in their $WV$ analysis CDF
performed a dedicated analysis of this shape. To focus on the
high-mass regime and to remove backgrounds they change some of the
cuts shown in Eq.(\ref{eq:cuts1}) to
\begin{alignat}{7}
\text{(1) exactly two jets with} \; 
E_{T,j} &> 30~\gev  
\qquad \qquad  \qquad \qquad 
\text{(2) additional dilepton veto} \; .
\label{eq:cuts2}
\end{alignat}
As we will see later, the veto on three or more jets makes a big
difference, both in the extraction of the signal and in the
uncertainties on the background estimates.  Unlike for the loose cuts
this experimental analysis show a distinct excess in
Fig.~\ref{fig:shapes_hard}. The additional requirements affects the
relative composition of all channels in the $m_{jj} = [28, 200]$~GeV
window~\cite{thesis}. For example, $WV$ production now contributes
6.4\% of all events, compared to 3.4\% for the loose cuts. The top
contribution very slightly decreases from 6.0\% to 5.8\%. For the two
mass windows we now find
\begin{alignat}{5}
\Delta N_{[64,96]}  
&=   \; 475 \;  \frac{\Delta \sigma_{WV} }{\sigma_{WV} }
&+&  \; 137 \;  \frac{\Delta \sigma_\text{top} }{\sigma_\text{top} }
\notag \\
\Delta N_{[120,170]} 
&=   \; 45 \; \frac{\Delta \sigma_{WV} }{ \sigma_{WV} }
&+&  \; 244 \; \frac{\Delta \sigma_\text{top} }{ \sigma_\text{top} }.
\label{eq:hard1}
\end{alignat}
Again, we use $\sigma$ for the cross section after cuts and
efficiencies, \ie $\sigma \times \epsilon_\text{cuts} \times
\epsilon_\text{rec}$. The relative normalization is fixed by the $WV$
peak region, giving us $(\Delta \sigma_{WV})/ \sigma_{WV} = -0.29 \,
(\Delta \sigma_\text{top})/\sigma_\text{top}$ and
\begin{equation}
\Delta N_{[120,170]} 
 = \; 231 \; \frac{\Delta \sigma_\text{top}}{ \sigma_\text{top}} \; .
\label{eq:hard2}
\end{equation}
Naively, we see around 230 events in the high mass region $m_{jj} =
120 - 170$~GeV. From this number we have to subtract the number of
events which are described by the $WV$ channel, including systematic
uncertainties. This leaves us with around 150 events which can for
example be explained by a Gaussian new physics contribution. 

However, this number of events changes after a more careful study of
the $m_{jj}$ distribution. First, in the $m_{jj} = 170 - 250$~GeV
range we see a significant tail, consistently 10 to 20 events above
the $WV$ expectations. They might be explained by some kind of
continuous background which would also contribute to the $m_{jj} = 170
- 250$~GeV window. Secondly, under the $WV$ peak of
Fig.~\ref{fig:shapes_hard} there are clearly events missing, of the
order of 50. Our simple compensation of the $WV$ and top
channels cannot account for them because they are missing in the left
side of the peak. Standard Model channels which rapidly drop towards
larger $m_{jj}$ values should help explaining them. This way we would
slightly decrease the number of events missing in the higher mass
regime.\medskip

Nevertheless, explaining an excess of more than 100 events in the
$m_{jj} = 120 - 170$~GeV requires a sizable shift in the
normalization of the top sample.  Eq.(\ref{eq:hard2}) implies $\Delta
\sigma_\text{top} \gtrsim 0.43 \, \sigma_\text{top}$ and a
compensating shift in the $WV$ rate of the order of
$\mathcal{O}(10\%)$.

Of course, this does not mean a $43\%$ shift in the theoretically
predicted total cross section for top production.  Almost a third of
the the combined top sample is single top production.  For the jet
veto survival probability the CDF analysis includes neither a reliable
experimental~\cite{cdf_single} nor a reliable theoretical
estimate~\cite{single_top}. Thus, we expect a very large error bar on
the single top rate after cuts and efficiencies.  Top pair production
might not be quite as critical because the parton shower approximation
should describe jets properly~\cite{skands,theo_scaling}.

All efficiencies very strongly depend on the detailed simulation of
the QCD jet activity and the $p_T$ requirements. For example, if we
increase the detection and veto threshold from 30~GeV to 40~GeV the
over-all efficiency increases quite dramatically for the top sample, as shown in
Fig.~\ref{fig:shapes_hard} and expected from Fig.~\ref{fig:peaks}. In
addition, it changes the shape of the top template. A reduced
efficiency for $WV$ events means that instead of Eq.(\ref{eq:hard2})
we find $(\Delta \sigma_{WV}/ \sigma_{WV}) = -0.68 \, (\Delta
\sigma_\text{top})/\sigma_\text{top}$ and makes it easier to explain
the second peak. This indicates large theory and systematic
uncertainties associated with the jet veto. The fact that it is
challenging to describe the top sample after jet related cuts is
illustrated by the poor separation of different single top channels in
the corresponding CDF analysis~\cite{cdf_single}. We check that the
corresponding uncertainty for loose cuts without a jet veto is very
well under control.\medskip

Taking our 40~GeV templates at face value the required change in the
combined top rate drops significantly, entirely due to a strong
dependence on the poorly understood jet veto survival probability. In
essence, subtracting combined top backgrounds after a jet veto
combines too many caveats which have to be taken into account as
correspondingly large systematic and theoretical
uncertainties\footnote{Very similar bottom lines will apply to many
  LHC searches to come.}.

\subsection*{Summary}
\label{sec:summary}

We have shown that the apparent excess in $W$+jet events can be
explained by Standard Model top backgrounds. Hadronically decaying top
quarks generically produce two peaks in the $m_{jj}$ distribution. To
explain the CDF measurements we have to enhance the normalization of
the combined top pair and single top templates after cuts and detector
efficiencies. Given the inherent difficulties in quantifying jet veto
survival probabilities, such a shift in the $10\%$ (for the $WW$
analysis without a jet veto) or the $40\%$ (for the high-mass analysis
with a jet veto) range appears reasonable and expected from QCD
considerations. To maintain the measured event numbers under the $WW$
peak we compensate for this shift in the top template with another
shift in the $WW$ normalization. The latter does not exceed 10\% and
is well within the uncertainties indicated by the different CDF
results for the individual electron and muon channels.\bigskip

Note added: after this work was finished, another paper with very
similar conclusions appeared~\cite{zack}.

\subsection*{Acknowledgments}

We are grateful for many discussions about the CDF anomaly here in
Heidelberg, including Michael Spannowsky, Steffen Schumann, Christoph
Englert, and Bob McElrath. Moreover, we would like to thank Tim Tait
for his comments on the manuscript and for pointing out
Ref.\cite{cdf_single}.


\end{document}

%% file: declare.tex
\newcommand\one{\leavevmode\hbox{\small1\normalsize\kern-.33em1}}

\newcommand{\met}{\slashchar{p}_T}

\newcommand{\gev}{{\ensuremath\rm GeV}}

\newcommand{\ifb}{{\ensuremath\rm fb^{-1}}}

\def\slashchar#1{\setbox0=\hbox{$#1$}           
   \dimen0=\wd0                                 
   \setbox1=\hbox{/} \dimen1=\wd1               
   \ifdim\dimen0>\dimen1                        
      \rlap{\hbox to \dimen0{\hfil/\hfil}}      
      #1                                        
   \else                                        
      \rlap{\hbox to \dimen1{\hfil$#1$\hfil}}   
      /                                         
   \fi}

\def\eg{{\sl e.g.} \,}
\def\ie{{\sl i.e.} \,}

\newcommand{\be}{\begin{eqnarray*}}
\newcommand{\ee}{\end{eqnarray*}}

\newcommand{\bee}{\begin{eqnarray}}
\newcommand{\eee}{\end{eqnarray}}
\newcommand{\beeq}{\begin{equation}}
\newcommand{\eeeq}{\end{equation}}